\documentclass[letterpaper,12pt]{article}

\usepackage{endnotes}
\usepackage{amssymb}
\usepackage{amsfonts}
\usepackage{amsmath}
\usepackage{amsthm}
\usepackage{graphicx,color}
 
\usepackage{comment}

\usepackage{array}

\usepackage{url}

\usepackage{setspace}
\def\R{\mathbb{R}}
\def\endproof{\hfill\diamondsuit}

\def\sF{{\mathcal F}}

\def\sA{{\mathcal A}}
\def\sL{{\mathcal L}}
\def\sM{{\mathcal M}}

\def\sS{{\mathcal S}}

\def\L{\mathbb{L}}

\def\E{\mathbb{E}}

\def\sF{\mathcal{F}}
\def\P{\mathbb{P}}

\def\Q{\mathbb{Q}}

\def\N{\mathbb N}

\newcommand{\Bigcdot}{\cdot}

\newcommand{\od}{\overline{\delta}}
\newcommand{\ud}{\underline{\delta}}

\newcommand{\norm}[1]{\Arrowvert #1 \Arrowvert}
\newcommand{\abs}[1]{\vert #1 \vert}

\numberwithin{equation}{section}
\theoremstyle{plain}                
\newtheorem{theorem}{Theorem}[section]
\newtheorem{lemma}[theorem]{Lemma}

\theoremstyle{definition}           
\newtheorem{definition}[theorem]{Definition}
\newtheorem{example}[theorem]{Example}

\newtheorem{assumption}[theorem]{Assumption}
\theoremstyle{remark}               

\usepackage[margin=1.25in]{geometry}

\begin{document}
\pagenumbering{arabic} \pagestyle{plain}

\begin{center}
\LARGE{\bf Taylor approximation of incomplete Radner equilibrium models}
\end{center}
\begin{center}
\ \\ \ \\
{\large \bf Jin Hyuk Choi}\\ Department of Mathematical Sciences, \\
  Carnegie Mellon University,\\ Pittsburgh, PA 15213, USA \\ email: {\tt
    jinhyuk@andrew.cmu.edu}

\ \\ 

{\large \bf Kasper Larsen}\\ Department of Mathematical Sciences, \\
  Carnegie Mellon University,\\ Pittsburgh, PA 15213, USA \\ email: {\tt
    kasperl@andrew.cmu.edu}

\end{center}
\ \\
\begin{center}

{\normalsize \today }
\end{center}

\ \\
\begin{verse}
{\sc Abstract}: In the setting of exponential investors and uncertainty governed by Brownian motions we first prove the existence of an incomplete equilibrium for a general class of models. We then introduce a tractable class of exponential-quadratic models and prove that the corresponding incomplete equilibrium is characterized by a coupled set of Riccati equations. Finally, we prove that these exponential-quadratic models can be used to approximate the incomplete models we studied in the first part.

\end{verse}
\vspace{0.5cm}

\newpage

\section{Introduction}

In a multi-dimensional auto-regressive Brownian setting with heterogeneous exponential utility investors we first prove that an incomplete equilibrium exists.  Each investor's endowment is allowed to contain  non tradable risk components which implies model incompleteness. Secondly, we construct a class of incomplete models for which the equilibrium is described by coupled Riccati equations. We then show that this tractable class of models can be used as a Taylor approximation of the general class of models we first considered. We construct an example showing that the established rate of convergence (seen as a function of the time-horizon) cannot be improved. 

Complete models, i.e., models in which the investors' income streams (endowments) can be traded, have been extensively studied in the literature and references include the textbooks \cite{KS98}, \cite{Duf01}, and \cite{DJ03}. Alternatively, when the investors' endowments cannot be traded, the underlying model is incomplete. The current literature on the existence of equilibria in  Brownian models with unspanned endowments and continuous-time trading is very limited. The theory related to incomplete equilibrium is significantly more involved mathematically because there is no simple a priori parameterization of all possible equilibria. For complete models, the representative agent provides such a parameterization via constant Pareto-efficient weights. The working paper \cite{CH94} generalizes the notion of a representative agent to include stochastic weights (non Pareto-efficient) needed to allow for model incompleteness. However, \cite{CH94} require certain properties of the dual optimizers which are hard to verify upfront (see Theorem 4 and 5 in \cite{CH94}). The paper \cite{CLM12} presents a model based on a multi-dimensional Brownian motion  which produces the incomplete equilibrium in closed-form and quantifies the negative effect income incompleteness can have on the equilibrium interest rate. The working paper \cite{CL12} extends \cite{CLM12} to include non traded stochastic income volatility components and shows that this feature can both lower the equilibrium interest rate and raise the equilibrium equity premium.

The paper \cite{Zit12} uses Banach's fixed point theorem to ensure the equilibrium existence in a model with noise generated by a single Brownian motion and an independent indicator process. The existence of an incomplete equilibrium  in the case of multiple Brownian motions is proven in the thesis \cite{Zha12} using Schauder's fixed point theorem under a decay property of the endowments (see Assumption 2.3.1 in \cite{Zha12}). We show how the proof in \cite{Zit12} can be adjusted to our setting where the underlying factor process is a multi-dimensional Ornstein-Uhlenbeck process driven by multiple Brownian motions. Furthermore, we remove the aforementioned decay property used in \cite{Zha12}. When compared to \cite{Zit12} and \cite{Zha12} our setting also includes an endogenously determined interest rate. 

In the second part of this paper we construct a class of exponential-quadractic models for which the corresponding equilibrium is characterized by a coupled set of Riccati equations. This class of models is highly tractable since it is characterized by simple coupled ordinary differential equations. 
We show that the general setting's  incomplete equilibrium can be approximated (for short time-horizons) by replacing the individual investors' endowments with their second degree Taylor approximations. This type of approximation falls into the setting of exponential-quadratic models. We show that the market price of risk process corresponding to the approximate equilibrium converges (in an $\L^1$-senese) to the market price of risk process corresponding to the general equilibrium as the time-horizon vanishes. We exemplify that the rate of convergence we have established cannot (in general) be improved.

The paper is organized as follows: In the next section we set up the model and the individual investors' problems. Section 3 contains the main existence result whereas in Section 4 we introduce the Taylor approximation and 
establish its convergence properties. All proofs are in the Appendix.
\section{Model setup}

\subsection{Mathematical setting and notation}
For a vector $x$, we denote by $x^T$ the transpose of $x$.
We let $\big(\Omega,\sF, \P\big)$ be a probability space on which $W= (W^{(1)},...,W^{(D)})^T$ is an $D$-dimensional Brownian motion, i.e., each coordinate process is a one dimensional Brownian motion and all coordinate processes are independent processes. We consider a unit time-horizon and we let the filtration $(\sF_t)_{t\in[0,1]}$ be the usual augmented filtration generated by $W$. For simplicity we will assume that $\sF = \sF_1$.

We briefly recall the following standard notation for stochastic integration, see, e.g., \cite{SC02}. For two $d$-dimensional processes $X$ and $Y$ with $X$ being a continuous semi-martingale we write $Y\in\sL_t(X)$ if $Y$ is a progressively measurable process for which the vector stochastic integral $\int_0^sY_udX_u=(Y\Bigcdot X)_s$ is well-defined for $s\in[0,t]$. 


\subsection{Factor process}
The underlying Markovian factor process will be denoted by $Y= (Y_t)_{t\in[0,1]}$ and is defined as follows: We consider deterministic, measurable, and locally bounded functions $A:[0,1]\to \R^D$ and $B,C:[0,1]\to \R^{D\times D}$. Then the following $D$-dimensional Ornstein-Uhlenbeck process is well-defined
\begin{align}\label{OU}
dY_t := \Big(A(t) + B(t) Y_t\Big)dt + C(t) dW_t,\quad Y_0 \in\R^D.
\end{align}
This choice of the underlying factor process has been widely used in the finance literature. In particular, the term structure models of Vasicek and Hull-White as well as their multi-factor extensions are based on such dynamics. We refer to \cite{DK96} for more details.

\subsection{Financial model}
We consider a pure-exchange-economy, i.e., there exists a single consumption good in which all prices are quoted. Investors can only consume initially ($t=0$) and at maturity ($t=T$) whereas trading can take place continuously throughout $[0,T]$ for $T\in(0,1]$. In addition to the money market account $S^{(0)}$, the investors can trade in $N$ non-dividend paying securities $S = (S^{(1)},...,S^{(N)})^T$. We always assume that $N\le D$ and whenever $N<D$ the resulting model is incomplete. 

In the next section we provide conditions under which the following assumption holds. The set $\sM$ denotes the set of equivalent local martingale measures $\Q$, i.e., $\Q\in\sM$ is a $\P$-equivalent probability measure under which $\tilde{S} := S/S^{(0)}$ has zero drift.
\begin{assumption} \label{conj_prices}
There exists a continuous function $\lambda:[0,T]\times\R^{D} \to \R^{N}$ and a constant $r\in\R$ such that:
\begin{enumerate} 

\item The money market account carries the constant interest rate $r$, i.e., its price process is $S^{(0)}_t =e^{rt}$. The dynamics of the price processes of the risky securities are well-defined as 
\begin{align}\label{dS}
dS^{(n)}_t = \big (\lambda(t,Y_t) + rS^{(n)}_t\big)dt + dW^{(n)}_t,\quad S_0^{(n)}=1,\quad n=1,2,...,N.
\end{align}

\item The process $\big(\lambda(t,Y_t))_{t\in[0,T]}$ ensures that $\sM$ is not empty.

\end{enumerate}

$\endproof$
\end{assumption}

In Assumption \ref{conj_prices} the interest rate is taken to be constant because  the investors can only consume initially and at maturity. Since the risky security is not paying dividends, its volatility structure and initial value remain undetermined. The diagonal volatility structure we use in \eqref{dS} has been chosen for its notational simplicity and can easily be replaced by a general stochastic volatility matrix. The $n$'th component of the $N$-dimensional vector $\lambda(t,Y_t)$ in \eqref{dS} is by definition the excess risk-free return of $S^{(n)}$ and is also called the market price of risk process for $W^{(n)}$. 

\subsection{Radner equilibrium}
Our model has $I\in\N$ heterogeneous exponential investors with coefficients $a_i>0$:
$$
U_i(b) := -e^{-a_i b},\quad b\in\R.
$$
We assume that each investor's subjective probability measure is $\P$.
We model the investors' endowments paid at time $T\in(0,1]$ by continuous functions $g^{(i)}:\R^D\to \R$ wheras the investors' initial endowments are denoted by $g_0^{(i)}\in\R$ for $i\in \{1,2,...,I\}$.

The note \cite{BS12} discusses several possible notions of admissibility when the utility function is defined on $\R$. We will use the following investor specific notion of admissibility.  We fix $i\in\{1,2,...,I\}$. Under Assumption \ref{conj_prices} and for a fixed measure $\hat{\Q}^{(i)}\in \sM$ we deem a process $H\in\sL_T^2(\tilde{S})$ admissible if $(H\Bigcdot \tilde{S})_t$ is a $\hat{\Q}^{(i)}$-supermartingale on $[0,T]$ in which case we write $H\in\sA_i= \sA_i\big(\hat{\Q}^{(i)}\big)$.
Investor $i$, $i=1,2,...,I,$ seeks $(\hat{c}^{(i)}_{0},\hat{H}^{(i)})\in\R\times\sA_i$ such that:
\begin{align}\label{ith_problem}
\begin{split}
\sup_{c_0\in\R,H\in\sA_i}\,& \E\left[U_i\left(c_0 + g^{(i)}_0\right)+U_i\left(X^{-c_0,H}_T + g^{(i)}\big(Y_T\big)\right)\right] \\&\quad= \E\left[U_i(\hat{c}^{(i)}_0 +g^{(i)}_0)+U_i\left(X^{-\hat{c}^{(i)}_0,\hat{H}^{(i)}}_T + g^{(i)}\big(Y_T\big)\right)\right],
\end{split}
\end{align}
where $dX_t^{x,H}:= rX^{x,H}_tdt + \sum_{n=1}^N(H_t)_n\big(\lambda(t,Y_t)_ndt  +dW^{(n)}_t\big)$ with $X^{x,H}_0:=x$. 

We adapt the following definition of a Radner equilibrium and for more information we refer to Chapter 5 in \cite{DJ03}.

\begin{definition}[Radner] \label{def:Radner} An equilibrium up to time $T\in(0,1]$ is a constant $r$ and a function $\lambda$ satisfying Assumption \ref{conj_prices} such that there exist measures $\hat{\Q}^{(i)}\in\sM$ and $(\hat{c}^{(i)}_0,\hat{H}^{(i)})\in \R\times\sA_i\big(\hat{\Q}^{(i)}\big)$ such that the pair $(\hat{c}^{(i)}_0,\hat{H}^{(i)})$ satisfies \eqref{ith_problem} for $i=1,2,...,I$, and the markets clear:
\begin{align}\label{clearing}
\sum_{i=1}^I \hat{c}^{(i)}_0 = 0,\quad \sum_{i=1}^I \hat{H}^{(i)}_t(\omega) = 0,\quad \text{for}\quad  \text{Leb}\otimes \P\text{-a.e.} \quad (t,\omega)\in[0,T]\times\Omega.
\end{align}
$\endproof$
\end{definition}

The self-financing property of the investors' wealth processes ensures that whenever \eqref{clearing} holds, the money market also clears, see, e.g., Remark 2.5 in \cite{Zit12}. Therefore, we will focus exclusively on \eqref{clearing} in what follows.


\subsection{Change of coordinates}
We will show that we can assume without loss of generality that $A=B=Y_0=0$ in \eqref{OU}. To see this, we let $\Phi:[0,1]\to \R^{D\times D}$ be the unique  solution of the following linear matrix equation
$$
\frac{\partial }{\partial t}\Phi(t) = B(t)\Phi(t),\quad \Phi(0) = I_{D\times D}, \quad t\in[0,1],
$$
where  $B$ is the matrix in \eqref{OU} and $I_{D\times D}$ is the $D\times D$-identity matrix. The unique solution $\Phi(t)$ exists and is non-singular for all $t\ge0$. We can then define the process
\begin{align*}
\tilde{Y}_t := \int_0^t\tilde{C}(s) dW_s,\quad \tilde{C}(t) := \Phi^{-1}(t) C(t),\quad t\in[0,1].
\end{align*}
The results in Section 5.5.6 in \cite{KS91} produce the representation
$g^{(i)}(Y_T) = \tilde{g}^{(i)}(\tilde{Y}_T)$ where
\begin{align}\label{tilde}
\tilde{g}^{(i)}(y):= g^{(i)}\left(\Phi(T)\Big(Y_0 + \int_0^T \Phi^{-1}(s) A(s) ds + y\Big)\right),\quad y\in\R^D.
\end{align}
This argument justifies the dynamics \eqref{OU1} in the following assumption. 
A discussion of H\"older spaces can be found in Appendix \ref{app:holder}.
\newpage
\begin{assumption}\label{ass:factor} There exist $\alpha\in(0,1)$ and $\overline{\delta} > \underline{\delta} >0$ such that the factor process $Y$ satisfies 
\begin{align}\label{OU1}
Y_t = \int_0^t C(u) dW_u,\quad t\in[0,1].
\end{align}
Here the function $C:[0,1]\to \R^{D\times D}$ satisfies that $C$ is $\alpha$-H\"older continuous and
\begin{align}\label{OU2}
\od |y|^2\ge y^T C(t)C(t)^T y \ge \ud|y|^2,\quad y\in\R^D,\quad t\in[0,1].
\end{align}
\end{assumption}
$\endproof$

\section{General existence result}

\begin{theorem}\label{thm:general} Under Assumption \ref{ass:factor}: We fix $\alpha \in (0,1)$ and we let $\left(g^{(i)}\right)_{i=1}^I \subset 
C^{2+\alpha}(\R^D)$. Then there exists $T_0\in(0,1]$ such that for all $T<T_0$ there exists an equilibrium $(r,\lambda)$ in the sense of Definition \ref{def:Radner}.
 \end{theorem}

This result extends Theorem 6.3.1 in \cite{Zha12} in several directions: Firstly, the decay property of Assumption 2.3.1 in \cite{Zha12} is not needed in Theorem \ref{thm:general}. Secondly,  as discussed in Section 2.5, Assumption \ref{ass:factor} allows for auto-regressivity in the underlying factor process. Finally,  Theorem \ref{thm:general} includes an equilibrium interest rate component.

From the proof Theorem \ref{thm:general} we see that there exists a constant $const>0$ which depends only on $(\ud,\od,D,\alpha,a_i,I,N)$ such that
$$
T_0\ge \frac{const}{\max_{i=1,...,I} |g^{(i)}|^2_{1+\alpha}}.
$$
In other words, large endowment functions (measured by the H\"older norms) produce smaller guaranteed valid maturities.

\section{Approximation}

We first introduce the highly tractable class of exponential-quadractic models. Then we show that these models can serve as second degree Taylor approximations of the general class of incomplete models we considered in Section 3. 

\subsection{Exponential-quadractic models}
We define the endowment functions $g^{(i)} :\R^D\to \R$ appearing in the optimization problems \eqref{ith_problem} by the quadratic form
\begin{align}\label{g_taylor}
g^{(i)}(y) := f^{(i)} + (h^{(i)})^T y + \frac12y^Tj^{(i)}y,\quad y\in\R^D,
\end{align}
where $f^{(i)} \in \R$, $h^{(i)} \in \R^D$, and $j^{(i)} \in \R^{D\times D}$. The proof of the next theorem shows that the Radner equilibrium $(r,\lambda)$ corresponding the endowment functions \eqref{g_taylor} can be characterized by a system of matrix-valued second order coupled ODEs. Consequently, the question of existence of an equilibrium can be reduced to ensuring the existence of a solution to a coupled system of Riccati equations. In the next theorem, the $D\times N$-matrix $\bar{C}(t)$ is defined by letting $\bar{C}(t)_{ij}$ denote $C(t)_{ij}$ for $i=1,...,D$ and $j=1,...,N$.

\begin{theorem}\label{thm:taylor} Let $Y$ be defined by \eqref{OU1} and let $g^{(i)}$ in \eqref{ith_problem} be defined by \eqref{g_taylor} for $i=1,2,...,I$. Then there exists a constant $T^\text{Riccati}_0\in(0,\infty]$ such that for all maturities $T<T^\text{Riccati}_0$ we have the following:  There exist a constant $r$ and continuous functions $\beta^{(i)}:[0,T]\to\R^D$, $\gamma^{(i)}:[0,T]\to \R^{D\times D}$, $i=1,2,...,I$,  such that the function 
\begin{align}\label{Taylor_mpr} 
\lambda(t,y):= \frac1{\tau_\Sigma} \bar{C}(t)^T \sum_{i=1}^I \Big(\beta^{(i)}(T-t) + \big(\gamma^{(i)}(t) + \gamma^{(i)}(T-t)^T\big)y\Big),\quad t\in [0,T],
\end{align}
together with $r$ forms an equilibrium in the sense of Definition \ref{def:Radner}.
\end{theorem}
The set of ODEs characterizing $\beta_i$ and $\gamma_i$ is provided in the proof of Theorem \ref{thm:taylor} (see Appendix A.4). 

\subsection{Taylor approximation}

The following result shows that the market price of risk process stemming from approximating $\big(g^{(i)}\big)_{i=1}^I$ with their second order Taylor approximation can be used to approximate the market price of risk process from the original model. We define
\begin{align}\label{gtilde}
\begin{split}
\tilde{g}^{(i)}&(y):=  g^{(i)}(0) +\big(\partial_y g^{(i)}(0)\big)^T y+ \frac12 y^T \partial_{yy} g^{(i)}(0)y, \quad y\in\R^D,
\end{split}
\end{align}
for $i=1,2,...,I$. This functional form is covered in the previous section and Theorem \ref{thm:taylor} produces the corresponding equilibrium. 
\begin{theorem}\label{thm:convergence} Under Assumption \ref{ass:factor}: For $\alpha\in(0,1)$ we let $\left(g^{(i)}\right)_{i=1}^I\subset C^{2+\alpha}(\R^D)$ and let $\lambda$ be the corresponding equilibrium market price of risk function produced by Theorem \ref{thm:general}. We let the functions $\tilde{g}^{(i)}$ be defined by \eqref{gtilde}  with corresponding equilibrium market price of risk function $\tilde{\lambda}$ produced by Theorem \ref{thm:taylor}. Then
\begin{align}\label{conv_rate}
\E\left[ \big|\lambda\big(t,Y_t)\big) - \tilde{\lambda}\big(t,Y_t\big)\big|\right] \le \text{const} \;T^{\frac{1+\alpha}2},\quad t\in[0,T],
\end{align}
where the constant \emph{const} is independent of both $t$ and $T$ and $T<T_0\land T_0^\text{Riccati}$ where $T_0>0$ is the maturity from Theorem \ref{thm:general} and $T^\text{Riccati}_0>0$ is the maturity from Theorem \ref{thm:taylor}.
\end{theorem}

 The convergence rate \eqref{conv_rate} is only valid for maturities $T\in (0,T_0\land T_0^\text{Riccati})$. Consequently,  the Taylor approximation is only guaranteed to work for short time-horizons. Furthermore, it can happen that $T_0 >T_0^\text{Riccati}$ in which case the approximate equilibrium exists on a shorter interval than the original equilibrium. If $T_0 >T_0^\text{Riccati}$ and we wish to approximate the incomplete equilibrium up to time $T_0$, we can replace  the second order Taylor approximation \eqref{gtilde} with its simpler first order analogue:
$$
\tilde{g}^{(i)}(y) := g^{(i)}(0) +\big(\partial_y g^{(i)}(0)\big)^T y,\quad y
\in\R^D.
$$
In this case, the incomplete equilibrium corresponding to the first order Taylor approximation exists on the full time-horizon $[0,1]$. The approximate market price of risk process becomes deterministic and the conclusion of Theorem \ref{thm:convergence} remains valid with the exponent $\frac{1+\alpha}2$ in \eqref{conv_rate} replaced by $\frac12$. 

We conclude this section with an example showing that the rate of convergence \eqref{conv_rate} established in Theorem \ref{thm:convergence} is in general optimal.

\begin{example} Let $\alpha \in (0,1)$ be fixed. We consider a single agent model,  i.e., $I:=1$, with risk aversion coefficient $a_1:=1$, and the complete model $Y_t := W^{(1)}_t$. We define the function
\begin{equation}
\begin{split}
f(x):=\left \{ \begin{array}{llll} 
2-|x|^{1+\alpha}, & |x|\le1, \\
(2-|x|)^{1+\alpha}, & |x|\in(1,2), \\
0, &\text{else}.
\end{array} \right.
\end{split}
\end{equation}
We will also need the function $F(x) := \int_{-2}^x f(y)dy$ for $x>-2$ and $F(x):=0$ for $x\le -2$. We then have that $f \in C^{1+\alpha}(\mathbb{R})$, hence, $F \in C^{2+\alpha}(\mathbb{R})$. 

We have $\lambda = \partial_y u$ and the characterizing PDE with $g^{(1)}(y) =F(y)$ becomes
\begin{equation}\label{ex pde}
\begin{split}
&\partial_tu + \frac{1}{2}\partial_{yy}u - \frac{1}{2}(\partial_yu)^2=0, \\
&u(T,y)=F(y),
\end{split}
\end{equation}
see Theorem \ref{thm:general_PDE} in the appendix. The explicit solution of \eqref{ex pde} is given by \begin{equation}
\begin{split}
u(t,y)= -\ln \Big( \int_{\mathbb{R}} \frac{1}{\sqrt{2\pi (T-t)}} e^{-\frac{x^2}{2(T-t)}} e^{-F(y-x)}dx \Big),\quad b\in\R,
\end{split}
\end{equation}
see, e.g., Chapter 4.4.1a in \cite{Eva10}. The expression for $\lambda=\partial_y u$ reads
\begin{equation}\label{ex:lambda}
\begin{split}
\lambda(t,y)= \frac{\int_{\mathbb{R}} e^{-\frac{x^2}{2(T-t)}} e^{-F(y-x)} F'(y-x)\,dx }{ \int_{\mathbb{R}}e^{-\frac{x^2}{2(T-t)}} e^{-F(y-x)}dx}.
\end{split}
\end{equation}

In the approximating model  we replace $F$ in 
\eqref{ex pde} by 
$$
\tilde{F}(y) := F(0) + F'(0)y + \frac12F''(0)y^2 = 2+2y,\quad y\in\R.
$$
In this case, formula \eqref{ex:lambda} produces the corresponding market price of risk function $\tilde{\lambda}(t,y) = 2$.  For $t:=0$ the left-hand-side of  \eqref{conv_rate} becomes
\begin{align*}
|\lambda(0,0) - \tilde{\lambda}(0,0)| &= \frac{\int_{\mathbb{R}} e^{-\frac{x^2}{2T}} e^{-F(-x)} (2-f(-x))dx }{ \int_{\mathbb{R}}e^{-\frac{x^2}{2T}} e^{-F(-x)}dx}\\
&\ge \frac{1}{\sqrt{2\pi T}}\int_{\mathbb{R}} e^{-\frac{x^2}{2T}} e^{-F(-x)} (2-f(-x))dx \\
&\ge e^{-4}\frac{1}{\sqrt{2\pi T}}\int_0^1e^{-\frac{x^2}{2T}}  x^{1+\alpha}dx 
\end{align*}
 The first equality holds because $\lambda  \le 2 = \tilde{\lambda}$ and the Gaussian kernel integrates to one. The first inequality holds because $F$ is positive whereas the second inequality follows from the properties $|f| \le 2$ and $|F|\le 4$. The last expression has the required form $const\,T^{\frac{1+\alpha}2}$ for some constant $const>0$.

$\endproof$

\end{example}
\appendix
\section{Proofs}

For $x\in\R^d$ we denote by $x_j$ the $j$'th coordinate whereas $|x|$ denotes the usual Euclidian 2-norm. If $X\in \R^{d\times n}$ has an inverse $X^{-1}$ we denote by $X^{-T}$ the transpose of $X^{-1}$. We will use the letter $c$ to denote various constants depending only on $(\ud,\od,D,\alpha,a_i,I,N)$. If the constant also depends on some H\"older norms we will use the letter $C$. The constants $c$ and $C$ never depend on any time variable. We do not relabel $c$ and $C$ from line to line.

\subsection{H\"older spaces}\label{app:holder} In this section we will briefly recall the standard notation related to H\"older spaces of bounded continuous functions, see, e.g., \cite{Kry96}. We fix $\alpha \in (0,1)$ in what follows.  The norm $|g|_0$ and the semi-norm $[g]_\alpha$ are defined by
$$
|g|_0:= \sup_{x\in \R^D} |g(x)|,\quad [g]_\alpha := \sup_{x,y \in \R^D, \,x\neq y}\frac{|g(x) - g(y)|}{|x-y|^\alpha},\quad g\in C(\R^D).
$$
We denote by $\partial_y g$ the vector of $g$'s derivatives and $\partial_{yy}g$ denotes the matrix of $g$'s second order derivatives. The H\"older norms are defined by
\begin{align*}
&|g|_\alpha:= |g|_0+[g]_\alpha, \quad g \in  C(\R^D),\\
&|g|_{1+\alpha}:= |g|_0+|\partial_y g|_0+[\partial_y g]_\alpha, \quad g \in  C^1(\R^D),\\
&|g|_{2+\alpha}:= |g|_0+|\partial_y g|_0+|\partial_{yy} g|_0+ [\partial_{yy}g]_\alpha, \quad g \in  C^2(\R^D),
\end{align*}
and the corresponding H\"older spaces are denoted by $C^{k+\alpha}(\R^D)$ for $k=0,1,2$. In these expressions we sum whenever the involved quantity is a vector or a matrix. So e.g., $|\partial_y g|_0$ denotes $\sum_{d=1}^D|\partial_{y_d} g|_0$ for a function $g=g(y) \in C^1(\R^D)$.

We also need the parabolic H\"older spaces for functions of both time and state. For such functions the usual supremum norm is defined by
$$
|u|_0:= \sup_{(t,x)\in [0,T]\times\R^D} |u(t,x)|,\quad u\in C([0,T]\times \R^D).
$$
We denote by $\partial_tu$ the partial derivative with respect to time of a function $u = u(t,x)$. The parabolic versions of the above H\"older norms are defined as
\begin{align*}
 |u|_{\alpha}&:= |u|_0 +[ u]_\alpha,\quad u\in C([0,T]\times \R^D),\\
 |u|_{1+\alpha}&:= |u|_0+|\partial_yu|_0+ [\partial_y u]_\alpha,\quad u\in C^{0,1}([0,T]\times \R^D),\\
 |u|_{2+\alpha}&:= |\partial_tu|_0 + [\partial_t u]_\alpha+ |u|_0+|\partial_y u|_0+|\partial_{yy} u|_0  + [\partial_{yy}u]_\alpha,\quad u\in C^{1,2}([0,T]\times \R^D),
\end{align*}
where  $\partial_y u$ and $\partial_{yy} u$ denote the first and second order derivatives with respect to the state variable and
$$
[h]_\alpha := \sup_{(t,x),(s,y) \in [0,T]\times\R^D, \\ (t,x)\neq (s,y)}\frac{|h(t,x) - h(s,y)|}{(\sqrt{|t-s|}+|x-y|)^\alpha},\quad h \in\{ \partial_y u, \partial_{yy}u, \partial_tu\}.
$$
The corresponding parabolic H\"older spaces are denoted by $C^{k+\alpha}([0,T]\times \R^D)$ for $k=0,1,2$.

We conclude this section with a simple inequality which we will need later.
\begin{lemma} \label{lem:simple_ineq}For $h_1,h_2, \tilde{h}_1$ and $\tilde{h}_2$ in $C^\alpha([0,T]\times \R^D)$  we have:
\begin{align}\label{simple_ineq}
\abs{h_1h_2 - \tilde{h}_1 \tilde{h}_2}_{\alpha} 
&\leq \frac{1}{2}\Big(\abs{h_1-\tilde{h}_1}_{\alpha} \, \abs{h_2+\tilde{h}_2}_{\alpha} +\abs{h_1+\tilde{h}_1}_{\alpha} \, \abs{h_2-\tilde{h}_2}_{\alpha} \Big).
\end{align}
\end{lemma}

\proof  Equation (3.1.6) in \cite{Kry96}  produces for $h_1,h_2\in C^\alpha([0,T]\times \R^D)$ the inequality
$$
[h_1 h_2]_\alpha \leq \abs{h_1}_0 [h_2]_\alpha + [h_1]_{\alpha} \abs{h_2}_0.
$$
From this inequality and the definition of $|\cdot|_\alpha$ we get
\begin{align*}
|h_1 h_2|_\alpha = |h_1h_2|_0 + [h_1h_2]_\alpha\le |h_1|_0|h_2|_0 + [h_1h_2]_\alpha \le |h_1|_\alpha |h_2|_\alpha.
\end{align*}
Consequently, since 
$$
\abs{h_1h_2 - \tilde{h}_1 \tilde{h}_2}_{\alpha} 
= \frac{1}{2}\Big|(h_1-\tilde{h}_1)(h_2+\tilde{h}_2)+(h_1+\tilde{h}_1)(h_2-\tilde{h}_2)\Big|_{\alpha},
$$
the triangle inequality produces \eqref{simple_ineq}.

$\endproof$

\subsection{Estimates from Linear Algebra}
We start with the following result from linear algebra which we need the next section. For a $D\times D$ positive definite matrix $X$ we denote by $||X||_F$ the Frobenius norm, i.e.,  
$$
||X||_F^2 := \sum_{i,j=1}^D X_{ij}^2.
$$
We note that Cauchy-Schwartz's inequality holds: $|Xx| \le ||X||_F |x|$ for $x\in\R^D$.
\begin{lemma}\label{lem:LA} Let $C$ satisfy Assumption \ref{ass:factor}. We define the $D\times D$-matrix
\begin{align}\label{def:Sigma}
\Sigma(t,s) := \int_t^s C(u)C(u)^T du, \quad 0\le t < s \le 1.
\end{align}
\begin{enumerate}
\item[(1)] The function $\Sigma$ is symmetric, positive definite,  and satisfies:
$$
|\Sigma(t,s)_{ij}| \le \od(s-t), \quad \ud(s-t)\le \Sigma(t,s)_{ii},\quad i,j=1,...,D.
$$
\item[(2)] The inverse $\Sigma(t,s)^{-1}$ exists and is symmetric, positive definite, and satisfies
$$
\frac1{\od(s-t)} \le \Sigma(t,s)^{-1}_{ii} \le \frac{1}{\ud(s-t)},\quad i=1,...,D.
$$
Consequently, $|\Sigma(t,s)_{ij}^{-1}| \le  \frac{1}{\ud(s-t)}$ for  $i,j=1,...,D$.
\item[(3)] The lower triangular matrix $L(t,s)$ in the Cholesky decomposition $\Sigma(t,s) = L(t,s)L(t,s)^T$ satisfies
$$
|L(t,s)_{ij}| \le \sqrt{\od(s-t)},\quad L(t,s)_{ii} \ge \sqrt{\ud(s-t)},\quad i,j=1,...,D.
$$
\item[(4)] For $0\le t_1 < t_2 < s$ we have $|| L(t_1,s)-L(t_2,s)||_F \le c \sqrt{t_2-t_1}$ where $c$ is a constant depending only on $\ud,\od, D$.
\item[(5)] There exists a constant $c$, depending only on $\ud,\od, D$, such that 
for $i=1,...,D$:
$$
\Big| \sqrt{\Sigma(t_1,s)_{ii}^{-1}} - \sqrt{\Sigma(t_2,s)^{-1}_{ii}}\Big| \le c \min\left\{\frac1{\sqrt{s-t_2}}, \frac{t_2-t_1}{(s-t_2)^{\frac32}}\right\},\quad 0\le t_1 < t_2 < s.
$$ 
\end{enumerate}
\end{lemma}
\proof (1): The symmetry follows from \eqref{def:Sigma}. For $y\in \R^D$, Condition \eqref{OU2} of Assumption \ref{ass:factor} produces
\begin{align}\label{LA1}
\frac{y^T \Sigma(t,s) y}{|y|^2} = \int_t^s \frac{y^T C(u)C(u)^T y}{|y|^2}du \in [\ud(s-t),\od(s-t)].
\end{align}
Therefore, $\Sigma(t,s)$ is also positive definite. By letting $y$ be the $i$'th basis vector $e_i\in \R^D$ we see 
$$
\ud(s-t) \le \Sigma(t,s)_{ii} \le \od(s-t).
$$
Finally, the inequality $|\Sigma(t,s)_{ij}| \le \sqrt{\Sigma(s-t)_{ii}\Sigma(s-t)_{jj}}$, see Problem 7.1.P1 in \cite{HJ13},  produces (1).

(2): Because $\Sigma(t,s)^{-1}$ is positive definite, the eigenvalues of $\Sigma(t,s)^{-1}$ are the reciprocal of the eigenvalues of $\Sigma(t,s)$. 
The claimed inequalities then follow from part (1) and Problem 4.2.P3 in \cite{HJ13}. 
The last estimate follows from $|\Sigma(t,s)^{-1}_{ij}| \le \sqrt{\Sigma(s-t)^{-1}_{ii}\Sigma(s-t)^{-1}_{jj}}$.

(3): To see the first claim we note that $\Sigma(t,s) = L(t,s)L(t,s)^T$ and (1) produce
$$
\sum_{j=1}^i L(t,s)^2_{ij} = \Sigma(t,s)_{ii} \le \od(s-t).
$$
To see the second claim we use Corollary 3.5.6, Theorem 4.3.17, and Corollary 7.2.9 in \cite{HJ13} to see
$$
L(t,s)_{ii} = \sqrt{\frac{i\text{'th leading principal minor of } \Sigma(t,s)}{(i\text{-1)'th leading principal minor of } \Sigma(t,s)}} \ge \sqrt{\ud(s-t)}.
$$

(4): We prove this by induction. By part (1) we have
\begin{align*}
|L(t_1,s)_{11}- L(t_2,s)_{11}| &= \sqrt{\Sigma(t_1,s)_{11}}-\sqrt{\Sigma(t_2,s)_{11}}\\
 &\le \sqrt{\Sigma(t_1,s)_{11}- \Sigma(t_2,s)_{11}} = \sqrt{\Sigma(t_1,t_2)_{11}} \le \sqrt{\od (t_2-t_1)}.
\end{align*}
For the induction step we suppose there is a constant $c$ such that $|L(t_1,s)_{ij} - L(t_2,s)_{ij}| \le c \sqrt{t_2-t_1}$ for $j=1,...,k-1$ and $i=j,...,D$. For $j=i=k$ we have
\begin{align*}
&|L(t_1,s)_{kk}- L(t_2,s)_{kk}| \\&= \frac{|L(t_1,s)_{kk}^2- L(t_2,s)^2_{kk}|}{L(t_1,s)_{kk}+ L(t_2,s)_{kk}}\\
&\le \frac{1}{\sqrt{\ud(s-t_1)}+\sqrt{\ud(s-t_2)}}\Big| \Sigma(t_1,t_2)_{kk} -\sum_{j=1}^{k-1}\Big(L(t_1,s)_{kj}^2 - L(t_2,s)^2_{kj}\Big)\Big|\\
&\le \frac{1}{\sqrt{\ud(s-t_1)}+\sqrt{\ud(s-t_2)}}\Big( \od(t_2-t_1) +\sum_{j=1}^{k-1}|L(t_1,s)_{kj} - L(t_2,s)_{kj}||L(t_1,s)_{kj} + L(t_2,s)_{kj}|\Big)\\
&\le \frac{1}{\sqrt{\ud(s-t_1)}+\sqrt{\ud(s-t_2)}}\Big( \od(t_2-t_1) +2c(k-1) \sqrt{t_2-t_1}\sqrt{\od(s-t_1)}\Big).
\end{align*}
The first inequality follows from (3). The second inequality follows from (1). The last inequality follows from (3) and the induction hypothesis. The last term is bounded by $c \sqrt{t_2-t_1}$ for some constant $c$.

For $j=k$ and $i = k+1,...,D$ we can use $\Sigma = LL^T$ to obtain the representation 
$$
L(t,s)_{ik} = \frac{\Sigma(t,s)_{ik} - \sum_{j=1}^{k-1} L(t,s)_{kj}L(t,s)_{ij}}{L(t,s)_{kk}}, \quad 0\le t <s,
$$
and arguments similar to the previous diagonal case to obtain the upper bound. All in all, we have the Frobenius norm estimate $|| L(t_1,s)-L(t_2,s)||_F \le c \sqrt{t_2-t_1}$. 

(5): By using $\tfrac{\partial }{\partial t} \Sigma(t,s)^{-1} = -\Sigma(t,s)^{-1} \tfrac{\partial }{\partial t} \Sigma(t,s) \Sigma(t,s)^{-1}$ we see for $0\le t<s$:
\begin{align*}
\Big|\frac{\partial}{\partial t} \sqrt{\Sigma(t,s)_{ii}^{-1}}\Big| &= \frac12 \Big|\frac1{\sqrt{\Sigma(t,s)_{ii}^{-1}}} \Big( \Sigma(t,s)^{-1} C(t)C(t)^T\Sigma(t,s)^{-1}\Big)_{ii}\Big|\\& \le \frac12 \Big|\sqrt{\od(s-t)} \od  \Big( \Sigma(t,s)^{-1} \Sigma(t,s)^{-1}\Big)_{ii}\Big|.
\end{align*}
Therefore, (2) gives us the bound
\begin{align*}
\Big|\frac{\partial}{\partial t} \sqrt{\Sigma(t,s)_{ii}^{-1}}\Big| \le c (s-t)^{-3/2},
\end{align*}
for some constant $c$. The Mean-Value Theorem then produces
$$
\Big| \sqrt{\Sigma(t_1,s)_{ii}^{-1}} -  \sqrt{\Sigma(t_2,s)_{ii}^{-1}}\Big| \le c \frac{t_2-t_1}{(s-t_2)^{3/2}}.
$$
This inequality combined with (2) concludes the proof.

$\endproof$

\subsection{Regularity of the Heat equation}

We let $\Sigma(s,t)$ be defined by \eqref{def:Sigma} and we let $\Gamma$ denote the following $D$-dimensional (inhomogenuous) Gaussian kernel:
\begin{align}\label{Gamma}
\Gamma(t,s,y):= \frac{e^{-\frac12 y^T\Sigma(t,s)^{-1} y}}{(2\pi)^{D/2}\text{det}\big(\Sigma(t,s)\big)^{1/2} },\quad 0\le t<s\le T,\quad y\in\R^D.
\end{align}

\begin{lemma}\label{lem:derivative} For $f_0\in C^{\alpha}([0,T]\times \R^D)$ we have for all $t\in[0,T]$ and $y\in\R^D$:
\begin{align*}
\partial_{y_d} \int_t^T \int_{\R^D} \Gamma(t,s,x-y) f_0(s,x) dxds = -\int_t^T \int_{\R^D}  \Gamma_{y_d}(t,s,x-y) f_0(s,x) dxds,
\end{align*}
for $d=1,...,D.$
\end{lemma}

\proof

We first assume that $f_0$ is continuously differentiable with compact support. In that case, the Dominated Convergence Theorem and integration by parts produce the claim. For $f_0$ merely continuous and bounded we approximate as follows: We first fix $R>0$. Since 
both $\Gamma(t,\cdot,\cdot)$ and $\Gamma_{y_d}(t,\cdot,\cdot)$ are integrable over $[t,T]\times \R^D$ we can find $M_n>R$ such that
$$
 \int_t^T \int_{|x|\ge M_n-R} \Gamma(t,s,x)  dxds\le \frac1n,\quad \int_t^T \int_{|x|\ge M_n-R}| \Gamma_{y_d}(t,s,x) | dxds\le \frac1n,\quad n\in\N.
$$
For each $n\in\N$, the density of compactly supported functions allows us to find a continuously differentiable function $f_n$ with compact  support such that 
$$
|f_n|_0 \le |f_0|_0,\quad \sup_{|x|\le M_n, s\in[t,T]} |f_n(s,x)-f_0(s,x)| \le \frac1n.
$$
For $|y|\le R$ we have $\{x\in\R^D: |x+y| > M_n\} \subset \{x\in\R^D: |x| > M_n -R\}$, hence, 
\begin{align*}
&\int_t^T \int_{\R^D} \Gamma(t,s,x) \big| f_n(s,y+x)-f_0(s,y+x)\big| dx ds\\
&\le \int_t^T \int_{|x+y|\le M_n} \Gamma(t,s,x) \big| f_n(s,y+x)-f_0(s,y+x)\big| dx ds\\
&\quad \quad +\int_t^T \int_{|x|> M_n-R} \Gamma(t,s,x) \big| f_n(s,y+x)-f_0(s,y+x)\big| dx ds\le \frac Tn +\frac{2|f_0|_0}n.
\end{align*}
A similar estimate (also uniform in $y$)  is found by replacing $\Gamma$ with $\Gamma_{y_d}$. For $|y|\le R$ and $t\in[0,T]$ we define the functions
\begin{align*}
&g_n(t,y) := \int_t^T \int_{\R^D} \Gamma(t,s,x-y) f_n(s,x)dxds,\quad n=0,1,...,\\
& h(t,y) := -\int_t^T \int_{\R^D} \Gamma_{y_d}(t,s,x-y) f_0(s,x)dxds.
\end{align*}
Since $f_n$ has compact support, we have $\partial_{y_d} g_n = - \int_t^T \int_{\R^D}\Gamma_{y_d} f_ndxds$. Therefore, 
$$
0=\lim_{n\to\infty} \sup_{|y|\le R} |g_n(t,y) - g_0(t,y)| = \lim_{n\to\infty} \sup_{|y|\le R} |\partial_{y_d} g_n(t,y)- h(t,y)|.
$$
The Fundamental Theorem of Calculus produces for $|y|\le R$:
\begin{align*}
&g(t,y_1,...,y_d,...,y_D) - g(t,y_1,...,0,...,y_D)\\
&=\lim_{n\to\infty}g_n(t,y_1,...,y_d,...,y_D) - g_n(t,y_1,...,0,...,y_D)\\
&=\lim_{n\to\infty}\int_0^{y_d} \partial_{y_d}g_n(t,y_1,...,\xi,...,y_D)d\xi = \int_0^{y_d} h(t,y_1,...,\xi,...,y_D)d\xi.
\end{align*}
Since $\partial_{y_d} g_n$ is continuous and converges uniformly to $h$ (on $|y|\le R$) we know that $h$ is also continuous. We can then apply $\partial_{y_d}$ to obtain  $\partial_{y_d}g = h$. Since $R>0$ was arbitrary the claim follows.

$\endproof$

\begin{lemma}\label{norm estimates} Under Assumption \ref{ass:factor}: For $\alpha \in (0,1)$ and $T\in[0,1]$ we let $f\in C^{\alpha}([0,T]\times \R^D)$ and $g \in C^{2+\alpha}(\R^D)$ be given. Then there exists a constant $c = c(\ud,\od, \alpha, D)$ and a unique solution $u \in C^{2+\alpha}([0,T]\times\R^D)$ of 
\begin{equation}\label{heat eq}
\begin{split}
\left \{ \begin{array}{ll}
u_t + \tfrac{1}{2} tr(\partial_{yy} u^{(i)} CC^T) + f = 0,\\
u(T,y)=g(y), \end{array} \right.
\end{split}
\end{equation}
which satisfies:
\begin{align}\label{Heat_1plusalpha_estimate}
\abs{u}_{1+\alpha} \leq c\left(\abs{g}_{1+\alpha} + \sqrt{T} \abs{f}_{\alpha}\right).
\end{align}
\end{lemma}

\proof  
Theorem 5.1 in Chapter 4 in \cite{LSU68} ensures the existence of a unique $C^{2+\alpha}([0,T]\times \R^D)$ solution  $u$ of \eqref{heat eq}. From Section 5.7B in \cite{KS91} we get the Feynman-Kac representation:
\begin{equation}\label{u exp}
u(t,y)= \int_{\R^D} \Gamma(t,T,x-y) g(x) dx + \int_t^T \int_{\R^D} \Gamma(t,s,x-y)f(s, x) dxds.
\end{equation}
  
From the representation \eqref{u exp} we immediately obtain $|u(t,y)| \le \abs{g}_0 + (T-t) \abs{f}_0$ which provides the norm estimate
\begin{align}
\abs{u}_0 \leq \abs{g}_0 + T \abs{f}_0, \label{estimate_|u|_0}
\end{align}
Since $\Sigma(t,s)$ is positive definite there exists a unique Cholesky decomposition $\Sigma(t,s) = L(t,s)L(t,s)^T$ for a lower non-singular triangular matrix $L(t,s)$. Furthermore, \\$\Sigma(t,s)^{-1} = L(t,s)^{-T}L(t,s)^{-1}$. By using det$(L(t,s))^2=$det$(\Sigma(t,s))$ when changing variables we can re-write \eqref{u exp}  as
\begin{equation}\label{u exp1}
\begin{split}
u(t,y)= \int_{\R^D} & \frac{e^{-\frac12|z|^2}}{(2\pi)^{D/2}}  g(y-L(t,T)z) dz  + \int_t^T \int_{\R^D} \Gamma(t,s,x-y)f(s, x) dxds.
\end{split}
\end{equation}

Since $g\in C^{2+\alpha}$ we can apply the Dominated Convergence Theorem on the $g$-integral and we can apply Lemma \ref{lem:derivative} on the $f$-integral in \eqref{u exp1} to produce:
\begin{equation}\label{u_b exp}
\begin{split}
u_{y_d}(t,y)&= \int_{\R^D}  \frac{e^{-\frac12|z|^2}}{(2\pi)^{D/2}}  g_{y_d}(y-L(t,T)z) dz \\&- \int_t^T \int_{\R^D}  \frac{e^{-\frac12|z|^2}}{(2\pi)^{D/2}}  \Big(L(t,s)^{-T}z\Big)_d f\big(s,y-L(t,s)z\big)dzds,
\end{split}
\end{equation}
after substituting $ z= L(t,s)^{-1}(y-x)$ in the $f$-integral. Since $||L(t,s)^{-T}||^2_F =\text{tr}(\Sigma^{-1}(t,s))\le \tfrac{D}{\ud(s-t)}$ by Lemma \ref{lem:LA}(2), Cauchy-Schwartz's inequality produces
\begin{align*}
\abs{u_{y_d}(t,y)} &\leq  \abs{g_{y_d}}_0 + |f|_0\int_t^T \int_{\R^D}  \frac{e^{-\frac12|z|^2}}{(2\pi)^{D/2}}  \Big|\Big(L(t,s)^{-T}z\Big)_d \Big|dzds \\
&\leq  \abs{g_{y_d}}_0 + |f|_0\int_t^T \int_{\R^D} \frac{e^{-\frac12|z|^2}}{(2\pi)^{D/2}}  ||L(t,s)^{-T}||_F|z|dzds \\
&\leq  \abs{g_{y_d}}_0 + D|f|_0\int_t^T\frac1{\sqrt{\ud(s-t)}} \int_{\R^D} \frac{e^{-\frac12|z|^2}}{(2\pi)^{D/2}} |z| dzds.
\end{align*}
By computing the integrals we obtain the estimate 
\begin{align}
&\abs{u_{y_d}}_0 \leq \abs{g_{y_d}}_0 +  c\sqrt{T} \abs{f}_0\label{estimate_|u_b|_0}.
\end{align}

To estimate the semi norm $[\partial_y u]_\alpha$ we will provide four estimates which when combined produce the estimate. We start by fixing  $0<t_1<t_2<T$ and $y_1,y_2\in \R^D$.  For the first estimate we have 
\begin{align*}
&\Big| \int_{\R^D} \frac{e^{-\frac12|z|^2}}{(2\pi)^{D/2}} \Big(g_{y_d}(y_1-L(t_1,T)z)-g_{y_d}(y_2-L(t_2,T)z)\Big) dz\Big|\\
&\le [g_{y_d}]_\alpha\int_{\R^D} \frac{e^{-\frac12|z|^2}}{(2\pi)^{D/2}}  \Big|y_1-L(t_1,T)z-y_2+L(t_2,T)z\Big|^\alpha dz,\\
&\le [g_{y_d}]_\alpha\int_{\R^D} \frac{e^{-\frac12|z|^2}}{(2\pi)^{D/2}}  \Big(|y_1-y_2| +||L(t_1,T)-L(t_2,T)||_F|z|\Big)^\alpha dz\\
&\le [g_{y_d}]_\alpha\int_{\R^D} \frac{e^{-\frac12|z|^2}}{(2\pi)^{D/2}}  \Big(|y_1-y_2| +c|t_1-t_2|^{1/2}|z|\Big)^\alpha dz.
\end{align*}
The first equality is due to the interpolation inequality which ensures that $[g_{y_d}]_\alpha<\infty$, see, e.g., Theorem 3.2.1 in \cite{Kry96}. The second inequality uses Cauchy-Schwartz's inequality whereas the last inequality is from Lemma \ref{lem:LA}(4).

The second estimate reads
\begin{align*}
&\Big|\int_{t_1}^{t_2} \int_{\R^D} \frac{e^{-\frac12|z|^2}}{(2\pi)^{D/2}} \Big(L(t_1,s)^{-T}z\Big)_d f\big(s,y_1-L(t_1,s)z\big)dzds\Big|\\
&=\Big|\int_{t_1}^{t_2} \int_{\R^D}  \frac{e^{-\frac12|z|^2}}{(2\pi)^{D/2}} \Big(L(t_1,s)^{-T}z\Big)_d \Big(f\big(s,y_1-L(t_1,s)z\big)-f\big(t_1,y_1\big)\Big)dzds\Big|\\
&\le c [f]_\alpha \int_{t_1}^{t_2}\frac1{\sqrt{s-t_1}}  \int_{\R^D}  \frac{e^{-\frac12|z|^2}}{(2\pi)^{D/2}} |z| \Big(|s-t_1|^{1/2} + |s-t_1|^{1/2}|z|\Big)^\alpha dzds\\
&\le c [f]_\alpha |t_2-t_1|^{(\alpha+1)/2},
\end{align*}
where the first inequality is found as before. The third estimate is similar and reads
\begin{align*}
&\Big|\int_{t_2}^T \int_{\R^D} \frac{e^{-\frac12|z|^2}}{(2\pi)^{D/2}} \Big(L(t_1,s)^{-T}z\Big)_d \Big(f\big(s,y_1-L(t_1,s)z\big)-f\big(s,y_2-L(t_2,s)z\big)\Big)dzds\Big|\\
& \le c[f]_\alpha\int_{t_2}^T \frac1{\sqrt{s-t_1}} \int_{\R^D} \frac{e^{-\frac12|z|^2}}{(2\pi)^{D/2}} |z| \Big(|y_1-y_2|  +|t_2-t_1|^{1/2} |z| \Big)^\alpha dzds.
\end{align*}

For the fourth and last estimate we first consider the case $d=D$. By the triangular structure of $L^{-1}$ and $L^{-T}$ we have $\sqrt{\Sigma^{-1}_{DD}}=L^{-1}_{DD}=L^{-T}_{DD}$. This gives us
\begin{align*}
&\Big|\int_{t_2}^T \int_{\R^D} \frac{e^{-\frac12|z|^2}}{(2\pi)^{D/2}} \Big\{\Big(L(t_1,s)^{-T}z\Big)_D - \Big(L(t_2,s)^{-T}z\Big)_D\Big\} f\big(s,y_2-L(t_2,s)z\big)dzds\Big|\\
&=\Big|\int_{t_2}^T \int_{\R^D} \frac{e^{-\frac12|z|^2}}{(2\pi)^{D/2}} \Big\{
\sqrt{\Sigma(t_1,s)^{-1}_{DD}}-\sqrt{\Sigma(t_2,s)^{-1}_{DD}} \Big\}z_D
\\&\quad \quad \times
\Big( f\big(s,y_2-L(t_2,s)z\big)-f\big(t_2,y_2\big)\Big)dzds\Big|\\
&\le c [f]_\alpha\int_{t_2}^T \min\left\{\frac1{\sqrt{s-t_2}}, \frac{t_2-t_1}{(s-t_2)^{\frac32}}\right\} \int_{\R^D} \frac{e^{-\frac12|z|^2}}{(2\pi)^{D/2}}  |z_D| 
\\&\quad \quad \times\Big( |s-t_2|^{1/2}+ |s-t_2|^{1/2}|z|\Big)^\alpha dzds,
\end{align*}
where the inequality follows from Lemma \ref{lem:LA}(5). The case $d <D$ can be reduced to the case $d=D$ we just considered by performing the following substitution: We let $J$ be the $D\times D$-matrix obtained by interchanging the $d$'th and $D$'th rows of the $D\times D$-identity matrix and we let $\tilde{L}$ be the lower triangular matrix in the Cholesky factorization $J\Sigma J = \tilde{L}\tilde{L}^T$. For $z:= \tilde{L}^{-1}J(y-x)$ we  have
\begin{align*}
\Big(\Sigma(t,s)^{-1}(y-x)\Big)_d &= \Big(J\Sigma(t,s)^{-1}(y-x)\Big)_D\\
&= \Big(J\Sigma(t,s)^{-1}JJ(y-x)\Big)_D = \Big(\tilde{L}(t,s)^{-T}z\Big)_D,
\end{align*}
where we used that $JJ$ is the $D\times D$-identity matrix and $J\Sigma^{-1}J = \tilde{L}^{-T}\tilde{L}^{-1}$.

These four estimates together with the triangle inequality as well as the representation \eqref{u_b exp} produce the parabolic semi-norm estimate 
\begin{align}\label{estimate_[u_b]_alpha}
&[u_{y_d}]_{\alpha} \leq c \big(|g|_{1+\alpha} + \sqrt{T}|f|_\alpha\big).
\end{align}
Finally, by combining the three estimates \eqref{estimate_|u|_0}, \eqref{estimate_|u_b|_0}, and \eqref{estimate_[u_b]_alpha} and using $T\le 1$ we produce the parabolic norm estimate \eqref{Heat_1plusalpha_estimate}.

$\endproof$
\begin{theorem}\label{thm:general_PDE}  Under the Assumptions of Theorem \ref{thm:general}: There exists $T_0\in(0,1]$ such that for all $T<T_0$ the non-linear PDE-system in $u^{(i)}= u^{(i)}(t,y)$ for $i=1,2,...,I$:
\begin{align*}
\begin{cases}
\partial_t u^{(i)} +  \frac1{2a_i} |\lambda|^2 - \lambda^T\bar{C}^T \partial_y u^{(i)} + \frac{a_i}2\left( |\bar{C}^T\partial_y u^{(i)}|^2-|C^T\partial_y u^{(i)}|^2\right)+ \frac12\text{tr}\left(\partial_{yy}u^{(i)}CC^T\right)=0,\\
u^{(i)}(T,y) = g^{(i)}(y),
\end{cases}
\end{align*}
where $\bar{C}$ is as in Theorem \ref{thm:taylor} and the coupling function $\lambda$ is defined as
\begin{align}\label{general_mpr}
\lambda(t,y) := \frac1{\tau_\Sigma}\bar{C}(t)^T\sum_{j=1}^I\partial_y u^{(j)}(t,y),
\end{align}
has a unique solution $\left(u^{(i)}\right)_{i=1}^I \subset C^{2+\alpha}([0,T]\times\R^D)$. 
 \end{theorem}
 
\proof We define $\sS_T:=\big(C^{1+\alpha}([0,T] \times \R^D)\big)^I$ for $I\in\N$, as well as the norm:
\begin{align*}
\norm{v}_{\mathcal{S}_T}:=\max_{i\in\{1,2,...,I\}} \abs{v^{(i)}}_{1+\alpha},\quad v\in\sS_T.
\end{align*}
Since $\big(C^{1+\alpha}([0,T] \times \R^D),|\cdot|_{1+\alpha}\big)$ is Banach space we also have that $(\sS_T, ||\cdot||_{\sS_T})$ is a Banach space.

In the following we will use the notation from Lemma \ref{norm estimates}. For $i=1,...,I$ we define the $i$'th coordinate $\Pi^{(i)}$ of the map $\Pi : \sS_T \to \sS_T$ by
\begin{displaymath}
\begin{split}
\Pi^{(i)}(v)(t,y):=&\int_{\R^D} \Gamma(t,T,x-y) g^{(i)}(x)dx + \int_t^T \int_{\R^D} \Gamma(t,s,x-y)f^{(i)}(v)(s,x) dxds, 
\end{split}
\end{displaymath}
where $f^{(i)}: \sS_T \to C^{\alpha}([0,T] \times \R^D)$ is defined by
\begin{align}\label{f exp}
f^{(i)}(v)&:= \frac1{2a_i} |\lambda(v)|^2 - \lambda(v)^T\bar{C}^T \partial_y v^{(i)} + \frac{a_i}2\left( |\bar{C}^T\partial_y v^{(i)}|^2-|C^T\partial_y v^{(i)}|^2\right),\\
\lambda(v)& := \frac1{\tau_\Sigma}\bar{C}^T\sum_{j=1}^I\partial_y v^{(j)}\nonumber.
\end{align}
Based on Lemma \ref{lem:simple_ineq} we have for $v,\tilde{v}\in\sS_T$ the estimates:
\begin{equation}\label{f ineq}
\begin{split}
\abs{f^{(i)}(v)}_\alpha &\leq c \norm{v}_{\mathcal{S}_T}^2,\\
\abs{f^{(i)}(v)- f^{(i)}(\tilde{v})}_{\alpha} &\leq c (\norm{v}_{\mathcal{S}_T}+\norm{\tilde{v}}_{\mathcal{S}_T}) \norm{v-\tilde{v}}_{\mathcal{S}_T},
\end{split}
\end{equation}
for a constant $c$. By combining \eqref{Heat_1plusalpha_estimate} with \eqref{f ineq} we produce the estimates 
\begin{displaymath}
\begin{split}
\abs{\Pi^{(i)}(v)}_{1+\alpha} &\leq  c \big(\abs{g^{(i)}}_{1+\alpha} + \sqrt{T} \norm{v}_{\mathcal{S}_T}^2\big), \\
\abs{\Pi^{(i)}(v)-\Pi^{(i)}(\tilde{v})}_{1+\alpha} & \leq c\sqrt{T}  (\norm{v}_{\mathcal{S}_T}+\norm{\tilde{v}}_{\mathcal{S}_T}) \norm{v-\tilde{v}}_{\mathcal{S}_T}.
\end{split}
\end{displaymath}
Therefore, by the definition of $\Pi$, we obtain the estimates
\begin{equation}\label{Pi ineq}
\begin{split}
\norm{\Pi(v)}_{\mathcal{S}_T} &\leq  c \big(\max_{1\leq i\leq I}\abs{g^{(i)}}_{1+\alpha} + \sqrt{T} \norm{v}_{\mathcal{S}_T}^2\big),\\
\norm{\Pi(v)-\Pi(\tilde{v})}_{\mathcal{S}_T} & \leq c\sqrt{T}  (\norm{v}_{\mathcal{S}_T}+\norm{\tilde{v}}_{\mathcal{S}_T}) \norm{v-\tilde{v}}_{\mathcal{S}_T}.
\end{split}
\end{equation}

To ensure that $\Pi$ is a contraction map, we consider real numbers $R>0$ and $T_0\in (0,1]$ such that (these constants $R$ and $T_0$ exist) 
\begin{equation}\label{T ineq}
\begin{split}
&c\Big(\max_{1\leq i\leq I}\abs{g^{(i)}}_{1+\alpha}+\sqrt{T_0} R^2\Big) \leq R, \\
&2c  \sqrt{T_0} R \leq \frac{1}{2}.
\end{split}
\end{equation}
For $T\in (0,T_0]$ we define the $R$-ball $\mathcal{B}_T:=\{ v\in \mathcal{S}_T : \norm{v}_{\mathcal{S}_T} \leq R \}\subset \Big(C^{1+\alpha}([0,T]\times \R^D)\Big)^I$. The estimates \eqref{Pi ineq} and the parameter restrictions \eqref{T ineq} produce that $\Pi$ maps $\mathcal{B}_T$ to $\mathcal{B}_T$ and that $\Pi$ is contraction map on $\mathcal{B}_T$. Since the space $(\mathcal{S}_T,||\cdot||_{\sS_T})$ is complete, there exists a unique fixed point $u\in \mathcal{B}_T$ of the map $\Pi$. The fixed point property $\Pi(u) = u$ implies that $u^{(i)}$ is given by \eqref{u exp} with $f := f^{(i)}(u)\in C^\alpha([0,T]\times \R^D)$. By uniqueness, we  obtain $u^{(i)} \in C^{2+\alpha}([0,T]\times \R^D)$. Consequently, the functions  $\big(u^{(i)}\big)_{i=1}^I\subset C^{2+\alpha}([0,T]\times \R^D)$ solve the stated PDE-system.

$\endproof$

\subsection{Remaining proofs}

We denote by $\binom{I}{0}$ the $D\times N$ matrix whose upper $N$ rows are the identity matrix $I_{N\times N}$ whereas all remaining entries are zeros. 

\proof[Proof of Theorem \ref{thm:general}] We will use Theorem \ref{thm:general_PDE} and let $T<T_0$. We can then define the function $\lambda= \lambda(t,y)$ by \eqref{general_mpr} as well as the constant
\begin{align}\label{r}
r:= \frac1{\tau_\Sigma T}\sum_{i=1}^I \Big(u^{(i)}(0,0) - g^{(i)}_{0}\Big).
\end{align}
The proof is split into the following two steps:

\noindent{\bf Step 1:} For $i=1,...,I$ we define the $N$-dimensional process
\begin{align}\label{FOC_HJB}
\hat{H}^{(i)}_t:= \frac1{a_i e^{r(T-t)}}\Big(\lambda(t,Y_t)- a_i\bar{C}(t)^T\partial_y u^{(i)}(t,Y_t)\Big),\quad t\in[0,T],
\end{align}
where $\bar{C}(t)_{ij}$ denotes $C(t)_{ij}$ for $i=1,...,D$ and $j=1,...,N$ . We will show that $\hat{H}^{(i)}$ is admissible in some set $\sA_i=\sA_i(\hat{\Q}^{(i)})$ and attains the supremum in
\begin{align}\label{ith_problem_no_c}
\sup_{H\in\sA_i}\,& \E\left[U_i\left(X^{0,H}_T + g^{(i)}\big(Y_T\big)\right)\right].
\end{align}
We note that in \eqref{ith_problem_no_c} the initial wealth is irrelevant because of the exponential preference structure. We define the function $V^{(i)}(t,x,y) := -e^{-a_i\big(e^{r(T-t)}x+u^{(i)}(t,y)\big)}$ as well as  the process $d\hat{X}^{(i)}_t := r\hat{X}^{(i)}_tdt + (\hat{H}^{(i)}_t)^T \Big(\lambda(t,Y_t) dt +\binom{I}{0}^TdW_t\Big)$ with $\hat{X}^{(i)}_0:=0$. It\^o's lemma produces the dynamics of $V^{(i)} = V^{(i)}(t,\hat{X}^{(i)}_t,Y_t)$ to be
\begin{equation}\label{dV^{(i)}}
\begin{split}
dV^{(i)} &= \partial_x V^{(i)} (\hat{H}^{(i)})^T\tbinom{I}{0}^TdW_t + (\partial_y V^{(i)})^TC(t)dW_t
\\
&=-V^{(i)} \Big\{\Big(\lambda^T- a_i\big(\partial_y u^{(i)})^T\bar{C}\Big)\binom{I}{0}^T+a_i(\partial_y u^{(i)})^TC\Big\}dW_t.
\end{split}
\end{equation}
Since the functions $\partial_yu^{(i)}$ and $\lambda$ are bounded, we can use Novikov's condition to see that $V^{(i)}$ is indeed a $\P$-martingale. 

Because $V^{(i)}(t,\hat{X}^{(i)}_t,Y_t) $ is a martingale, we have $q^{(i)}:= \E\big[e^{rT}U_i'\big(\hat{X}^{(i)}_T+ g^{(i)}(Y_T)\big)\big]\in(0,\infty)$. We can then define the $\P$-equivalent probability measures $\hat{\Q}^{(i)}$ via the Radon-Nikodym derivatives on $\sF_T$:
$$
\frac{d\hat{\Q}^{(i)}}{d\P} := \frac{V^{(i)}\big(T,\hat{X}^{(i)}_T,Y_T\big)}{V^{(i)}(0,0,0)}= \frac{e^{rT}U_i'\big(\hat{X}^{(i)}_T+ g^{(i)}(Y_T)\big)}{q^{(i)}},\quad i=1,2,...,I,
$$
where the last equality follows from the terminal condition $u^{(i)}(T,y) = g^{(i)}(y)$. We will next prove that $\hat{\Q}^{(i)}\in\sM$.  By the martingale property of $V^{(i)}$ we have 
$$
\E\left[\frac{d\hat{\Q}^{(i)}}{d\P}\Big|\sF_t\right] =\frac{V^{(i)}(t,\hat{X}^{(i)}_t,Y_t)}{V^{(i)}(0,0,0)},\quad t\in[0,T]. 
$$
Therefore, the dynamics \eqref{dV^{(i)}} of $dV^{(i)}$ together with Girsanov's Theorem ensure that  
$\tilde{S}:= S/S^{(0)}$ is an $N$-dimensional $\hat{\Q}^{(i)}$-martingale, hence, $\hat{\Q}^{(i)}\in\sM$. Since $\tilde{S}$'s volatility is $e^{-rt}$ and the process $\hat{H}^{(i)}$ defined by \eqref{FOC_HJB} is uniformly bounded, we have that the process $\hat{X}^{(i)}_te^{-rt}$ is a $\hat{\Q}^{(i)}$-martingale for $t\in[0,T]$, hence, $\hat{H}^{(i)}\in\sA_i$.

Finally, the verification of $\hat{H}^{(i)}$'s optimality is fairly standard and can be seen as follows.  Fenchel's inequality produces $U_i(x) \le U^*_i(y) + xy$ for all $x\in\R$ and $y>0$ where $U^*_i$ is the convex conjugate of $U_i$, i.e., $U^*_i(y):= \sup_{x\in\R}\big( U_i(x) - xy\big)$. Therefore, for arbitrary $H\in\sA_i$, we have 
\begin{align*}
\E\Big[U_i&\big( X^{0,H}_T + g^{(i)}(Y_T)\big)\Big]\\&\le \E\Big[U^*_i\big(q^{(i)}\frac{d\hat{\Q}^{(i)}}{d\P}e^{-rT}\big)\Big] + q^{(i)}\E\Big[\frac{d\hat{\Q}^{(i)}}{d\P}e^{-rT}\big( X^{0,H}_T + g^{(i)}(Y_T)\big)\Big] \\
&\le \E\Big[U^*_i\big(q^{(i)}\frac{d\hat{\Q}^{(i)}}{d\P}e^{-rT}\big)\Big] +q^{(i)} \E\Big[\frac{d\hat{\Q}^{(i)}}{d\P}e^{-rT}g^{(i)}(Y_T)\Big] \\
&= \E\Big[U^*_i\big(q^{(i)}\frac{d\hat{\Q}^{(i)}}{d\P}e^{-rT}\big)\Big] +q^{(i)} \E\Big[\frac{d\hat{\Q}^{(i)}}{d\P}e^{-rT}\big( \hat{X}^{(i)}_T + g^{(i)}(Y_T)\big)\Big] \\
&=\E\Big[U_i\big( \hat{X}^{(i)}_T + g^{(i)}(Y_T)\big)\Big].
\end{align*}
The second inequality is produced by the $\hat{\Q}^{(i)}$-supermartingale property of $(H\Bigcdot \tilde{S})_t$, the first equality is produced by the $\hat{\Q}^{(i)}$-martingale property of $(\hat{H}^{(i)}\Bigcdot \tilde{S})_t$ and the last equality follows from the first order condition in the definition $U^*$, see, e.g., Lemma 4.3(i) in \cite{KS98}. This verifies that $\hat{H}^{(i)}$ attains the supremum in \eqref{ith_problem_no_c}. \ \\

\noindent{\bf Step 2:} Based on the previous step we can re-write the optimization problem \eqref{ith_problem} as
$$
\sup_{c_0\in \R} \Big(-e^{-a_i(c_0 + g_0)} -e^{a_ie^{rT}c_0 -a_i u^{(i)}(0,0)}\Big).
$$
It is straightforward to solve this problem for $\hat{c}^{(i)}_0$ and see that \eqref{r} ensures the clearing condition $\sum_{i=1}^I\hat{c}^{(i)}_0=0$. 

$\endproof$


\proof[Proof of Theorem \ref{thm:taylor}]  The functions $\alpha^{(i)}(t) \in \R, \;\beta^{(i)}(t) \in \R^D$ and $\gamma^{(i)}(t) \in \R^{D\times D}$ for $t\ge0$ and $i=1,2,...,I$ are determined by the coupled ODEs .
\begin{align*}
&(\gamma^{(i)})' = \frac{a_i}2(\gamma^{(i)} + (\gamma^{(i)})^T)(\bar{C}\bar{C}^T-CC^T )(\gamma^{(i)} + (\gamma^{(i)})^T)\\&-\frac1{\tau_\Sigma}(\gamma^{(i)}+(\gamma^{(i)})^T)\bar{C}\bar{C}^T\sum_{j=1}^I (\gamma^{(j)}+(\gamma^{(j)})^T)\\& +\frac1{2a_i\tau_\Sigma^2}\left(\sum_{j=1}^I(\gamma^{(j)}+(\gamma^{(j)})^T)\right)\bar{C}\bar{C}^T\left(\sum_{j=1}^I(\gamma^{(j)}+(\gamma^{(j)})^T)\right),\quad \gamma^{(i)}(0)=j^{(i)},\\
&(\beta^{(i)})' =a_i(\gamma^{(i)}+(\gamma^{(i)})^T)(\bar{C}\bar{C}^T-CC^T)\beta^{(i)}+\frac1{a_i\tau_\Sigma^2}\left(\sum_{j=1}^I(\gamma^{(j)}+(\gamma^{(j)})^T)\right)\bar{C}\bar{C}^T\sum_{j=1}^I\beta^{(j)}\\&-\frac1{\tau_\Sigma}\left(\sum_{j=1}^I(\gamma^{(j)}+(\gamma^{(j)})^T)\right) \bar{C}\bar{C}^T\beta^{(i)}-\frac1{\tau_\Sigma}(\gamma^{(i)}+(\gamma^{(i)})^T)\bar{C}\bar{C}^T\sum_{j=1}^I\beta^{(j)},\;
\beta^{(i)} (0) = h^{(i)},\\
&( \alpha^{(i)})'  =\frac12 \text{tr}\big( (\gamma^{(i)}+(\gamma^{(i)})^T)CC^T\big) +\frac1{2a_i\tau_\Sigma^2}\Big| \bar{C}^T\sum_{j=1}^I \beta^{(j)}\Big|^2\\
 &\quad -\frac1{\tau_\Sigma}\Big(\sum_{j=1}^I (\beta^{(j)})^T\Big)\bar{C}\bar{C}^T\beta^{(i)}-\frac{a_i}2\Big(|C^T\beta^{(i)}|^2 - |\bar{C}^T\beta^{(i)}|^2\Big) ,\quad \alpha^{(i)}(0) = f^{(i)}.
\end{align*}
Since the right-hand-side is locally Lipshitz continuous seen as a function of the left-hand-side, there exists a unique solution up to some explosion time $T^\text{Riccati}_0\in(0,\infty]$ by the Picard-Lindel\"of Theorem. For $i=1,2,...,I$ we consider the quadratic form
\begin{equation}\label{u_approx}
u^{(i)}(t,y) := \alpha^{(i)}(T-t) + \big(\beta^{(i)}(T-t)\big)^Ty + y^T \gamma^{(i)}(T-t) y,\quad t\in[0,T],\quad y\in\R^D.
\end{equation}
 By computing the various derivatives, we see that \eqref{u_approx} solves the coupled PDE system in Theorem \ref{thm:general_PDE}.

It remains to perform verification and here we will just point out how the proof of Theorem \ref{thm:general} can be adjusted to the present case where $g^{(i)}$ is a quadratic 
function. The first issue is the martingale property of the process $V^{(i)}$ with the dynamics \eqref{dV^{(i)}}. As before, the process $V^{(i)}(t,\hat{X}^{(i)}_t,Y_t)$ - with  
$u^{(i)}$ defined by \eqref{u_approx} and $V^{(i)}(t,x,y) := -e^{-a_i\big(e^{r(T-t)}x+u^{(i)}(t,y)\big)}$ - is a local martingale under $\P$ with the dynamics \eqref{dV^{(i)}}. To see that $V^{(i)}$ is a martingale, we note that the partial derivative $\partial_yu^{(i)}$ is an affine function of $y$ with deterministic continuous functions of time-to-maturity as coefficients. Therefore, the function $\lambda$ defined by \eqref{general_mpr} is also an affine function of $y$. Because $dY = C(t)dW_t$ we can use Corollary 3.5.16 in \cite{KS91} to see that $V^{(i)}$ is martingale on $[0,T]$ for  $T<T^\text{Riccati}_0$.

Secondly, we need to prove the $\hat{\Q}^{(i)}$-martingale property of $d(\hat{X}^{(i)}_te^{-rt})= (\hat{H}^{(i)}_t)^Td\tilde{S}_t$ with $\hat{\Q}^{(i)}$ defined via $V^{(i)}$ as in the proof of Theorem \ref{thm:general}. The dynamics \eqref{dV^{(i)}} and Girsanov's Theorem produce the $\hat{\Q}^{(i)}$-Brownian motions
\begin{equation}
\begin{split}
dW^{\hat{\Q}^{(i)},m}_t:=\left \{ \begin{array}{llll} 
dW_t^{(m)} + \lambda(t,Y_t)_mdt, & m=1,...,N, \\
dW_t^{(m)} + a_i\Big( \big(\partial_y u^{(i)}(t,Y_t)\big)^TC(t)\Big)_mdt, & m = N+1,...,D.
\end{array} \right.
\end{split}
\end{equation}
Therefore, the drift of the $\hat{\Q}^{(i)}$-dynamics of $dY_t = C(t)dW_t$ is an affine function of $Y_t$ with bounded time-dependent coefficients.  

Since $\tilde{S}$'s volatility is $e^{-rt}$, it suffices to verify the square integrability property $\E^{\hat{\Q}^{(i)}}\left[\int_0^T \big(\hat{H}^{(i)}_t\big)^2dt\right]=\int_0^T \E^{\hat{\Q}^{(i)}}\left[\big(\hat{H}^{(i)}_t\big)^2\right]dt<\infty$ where $\hat{H}^{(i)}$ is defined by \eqref{FOC_HJB}. To this end, we define the stopping times:
$$
\tau^{(k)} := \inf\{s>0:|Y_s|\ge k\}\land T,\quad \text{for}\quad k\in\N.
$$
Because the functions $\lambda$ and $\partial_y u^{(i)}$ are affine with uniformly bounded time-dependent coefficients, the above expression for $W_t^{\hat{\Q}^{(i)},m}$ allows us to find two positive constants $C_1$ and $C_2$ (independent of $k$) such that
\begin{align*}
\E^{\hat{\Q}^{(i)}}\left[|Y_{t\land \tau^{(k)}}|^2\right] &\le C_1 + C_2\E^{\hat{\Q}^{(i)}}\left[\int_0^{t\land \tau^{(k)}} |Y_s|^2ds \right] \\
&= C_1 + C_2\E^{\hat{\Q}^{(i)}}\left[\int_0^{t\land \tau^{(k)}} |Y_{s\land \tau^{(k)}}|^2ds \right] \\
&\le C_1 + C_2\int_0^t \E^{\hat{\Q}^{(i)}}\left[|Y_{s\land \tau^{(k)}}|^2\right] ds,
\end{align*}
where we have used Tonelli's Theorem in the last inequality. The map $[0,T]\ni t\to \E^{\hat{\Q}^{(i)}}\left[|Y_{t\land \tau^{(k)}}|^2\right]$ is continuous by the Dominated Convergence Theorem. Therefore, Gronwall's inequality produces the bound
$$
\E^{\hat{\Q}^{(i)}}\left[|Y_{t\land \tau^{(k)}}|^2\right] \le C_1e^{C_2t},\quad t\in[0,T].
$$
Fatou's Lemma then produces
$$
\E^{\hat{\Q}^{(i)}}\left[|Y_t|^2\right]  \le \liminf_{n\to\infty} \E^{\hat{\Q}^{(i)}}\left[|Y_{t\land \tau^{(k)}}|^2\right] \le C_1e^{C_2t},\quad t\in[0,T].
$$
Finally, the definition \eqref{FOC_HJB} of $\hat{H}^{(i)}$ and the affinity of $\lambda$ and $\partial_y u^{(i)}$  ensure that there exists a constant $C_3$ such that $\E^{\hat{\Q}^{(i)}}\left[\big(\hat{H}^{(i)}_t\big)^2\right]\le C_3e^{C_2t}$. This latter expression is integrable on $[0,T]$ and the claim follows.
 

$\endproof$

\proof[Proof of Theorem \ref{thm:convergence}] In this proof, the functions $\tilde{\lambda}, \tilde{u}$ and $\partial_y\tilde{u}$ refer to the functions from Theorem \ref{thm:taylor} (and its proof) when the endowment functions \eqref{g_taylor} are specified by Taylor approximations \eqref{gtilde}. 

Definitions \eqref{general_mpr} and  \eqref{Taylor_mpr} of $\lambda$ and $\tilde{\lambda}$ together with the triangle inequality produce 
\begin{align*}
\mathbb{E}\Big[ &\big\vert \lambda(t,Y_t)- \tilde{\lambda}(t,Y_t)\big\vert \Big] 
\le \frac{1}{\tau_\Sigma}  \bar{C}(t)^T \sum_{i=1}^I  \mathbb{E}\Big[ \Big|\partial_y u^{(i)} (t,Y_t) -\partial_y \tilde{u}^{(i)} (t,Y_t) \Big| \Big].
\end{align*}
Therefore, it is enough to show that 
\begin{equation}\label{estimate0}
\begin{split}
\mathbb{E}\Big[ \Big|\partial_{y_d} u^{(i)} (t,Y_t) -\partial_{y_d} \tilde{u}^{(i)} (t,Y_t) \Big| \Big] \leq CT^{\frac{1+\alpha}{2}},\quad d=1,...,D,\quad i=1,...,I.
\end{split}
\end{equation}

From the representation \eqref{u_b exp} of $\partial_yu^{(i)}$ with $f$ replaced by $f^{(i)}(u)$, where $f^{(i)}(u)$ is defined by \eqref{f exp}, we get
\begin{align}\label{Taylor1}
&\abs{\partial_{y_d}u^{(i)}(t,y)-\partial_{y_d}g^{(i)}(y) } \nonumber \\
\begin{split}
&\leq \Big| \int_{\R^D}  \frac{e^{-\frac12|z|^2}}{(2\pi)^{D/2}}  \Big(g^{(i)}_{y_d}(y-L(t,T)z)   -g^{(i)}_{y_d}(y)\Big) dz \Big\vert \\
&\quad \quad + \Big\vert \int_t^T \int_{\R^D}  \frac{e^{-\frac12|z|^2}}{(2\pi)^{D/2}}  \Big(L(t,s)^{-T}z\Big)_d f^{(i)}(u)\big(s,y-L(t,s)z\big)dzds\Big\vert.\\
\end{split}
\end{align}
The first term in \eqref{Taylor1} can be estimated as follows:
\begin{displaymath}
\begin{split}
\Big\vert\int_{\R^D}  &\frac{e^{-\frac12|z|^2}}{(2\pi)^{D/2}}  \Big(g^{(i)}_{y_d}(y-L(t,T)z)   -g^{(i)}_{y_d}(y)\Big) dz \Big\vert  \\
&=\Big\vert\int_{\R^D}  \frac{e^{-\frac12|z|^2}}{(2\pi)^{D/2}} g_{yy_d}^{(i)}(s)^T L(t,T)z dz \Big\vert \\
&=\Big\vert\int_{\R^D}  \frac{e^{-\frac12|z|^2}}{(2\pi)^{D/2}}\Big( g_{yy_d}^{(i)}(s)^T-g_{yy_d}^{(i)}(y)^T  \Big)L(t,T)z dz \Big\vert \\
&\leq [\partial_{yy}g^{(i)}]_\alpha \int_{\R^D}  \frac{e^{-\frac12|z|^2}}{(2\pi)^{D/2}} |s-y|^\alpha  |L(t,T)z| dz  \\
&\leq [\partial_{yy}g^{(i)}]_\alpha \int_{\R^D}  \frac{e^{-\frac12|z|^2}}{(2\pi)^{D/2}} |L(t,T)z|^\alpha  |L(t,T)z| dz  \\
&\leq c[\partial_{yy}g^{(i)}]_\alpha  (T-t)^{(1+\alpha)/2}\int_{\R^D}  \frac{e^{-\frac12|z|^2}}{(2\pi)^{D/2}}   |z|^{\alpha+1}dz  \\
&= c [\partial_{yy}g^{(i)}]_\alpha (T-t)^{(1+\alpha)/2}.
\end{split}
\end{displaymath}
The first equality is produced by the Mean Value Theorem where $s = s(z)$  is on the line segment connecting $y-L(t,T)z$ and $y$. The third and fourth inequality are due to Cauchy-Schwartz: $|L(t,T)z| \le || L(t,T)||_F |z|$ combined with Lemma \ref{lem:LA}(3).

The second term in \eqref{Taylor1} can be estimated similarly: 
\begin{displaymath}
\begin{split}
&\Big\vert \int_{\R^D}  \frac{e^{-\frac12|z|^2}}{(2\pi)^{D/2}}  \Big(L(t,s)^{-T}z\Big)_d f^{(i)}(u)\big(s,y-L(t,s)z\big)dz\Big\vert \\
&=\Big\vert \int_{\R^D}  \frac{e^{-\frac12|z|^2}}{(2\pi)^{D/2}}  \Big(L(t,s)^{-T}z\Big)_d \Big(f^{(i)}(u)\big(s,y-L(t,s)z\big)-f^{(i)}(u)\big(s,y\big)\Big)dz\Big\vert \\
&\leq  [f^{(i)}(u)]_{\alpha} \int_{\R^D}  \frac{e^{-\frac12|z|^2}}{(2\pi)^{D/2}}   ||L(t,s)^{-T}||_F|z| |L(t,s)z|^\alpha dz\\
&\leq c [f^{(i)}(u)]_{\alpha} (s-t)^{\frac{\alpha-1}{2}}.
\end{split}
\end{displaymath}
By integrating $s$ over $[t,T]$ we produce the overall estimate of \eqref{Taylor1}:  
\begin{equation}\label{estimate1}
\begin{split}
\abs{\partial_{y_d}u^{(i)}(t,y)-\partial_{y_d}g^{(i)}(y) } \leq  C (T-t)^{(1+\alpha)/2}.
\end{split}
\end{equation}
We also have
\begin{align}
\abs{\partial_{y_d}g^{(i)}(y)-\partial_{y_d}\tilde{g}^{(i)}(y)} 
&= \abs{\partial_{y_d}g^{(i)}(y)-\partial_{y_d}g^{(i)}(0)-\partial_{yy_d}g^{(i)}(0)^Ty}\nonumber \\
&= \abs{\partial_{yy_d}g^{(i)}(s)^Ty -\partial_{yy_d}g^{(i)}(0)^Ty}\nonumber \\
&\leq \abs{\partial_{yy_d}g^{(i)}(s)-\partial_{yy_d}g^{(i)}(0)}|y| \nonumber\\
&\leq [\partial_{yy} g]_\alpha \abs{y}^{1+\alpha}\label{estimate2}.
\end{align}
Here the first equality follows form the definition of $\tilde{g}^{(i)}$. The second equality is produced by the Mean Value Theorem for a point $s = s(y)$ on the line segment connecting $y$ and $0$. Finally, we claim that there exists a constant $C$ such that
\begin{align}
\abs{\partial_{y_d}\tilde{g}^{(i)}(y) - \partial_{y_d}\tilde{u}^{(i)} (t,y)} 
&\leq C (T-t) (1+\abs{y})\label{estimate3}.
\end{align}
To see this, we first note that
\begin{align*}
\abs{\partial_{y_d}\tilde{g}^{(i)}(y) &- \partial_{y_d}\tilde{u}^{(i)} (t,y)} \\
&= \Big|\partial_{y_d}g^{(i)}(0)+\partial_{yy_d} g^{(i)}(0)^T y - \beta^{(i)}(T-t)_d -\Big(\big(\gamma^{(i)}(T-t) +\gamma^{(i)}(T-t)^T\big)y\Big)_d\Big|.
\end{align*}
Since $ \beta^{(i)}(0)_d = \partial_{y_d}g^{(i)}(0)$ the Mean Value Theorem gives us
$s \in [0,T-t]$ such that
\begin{align*}
\abs{\partial_{y_d}g^{(i)}(0) - \beta^{(i)}(T-t)_d } = \abs{(\beta^{(i)})'(s)_d(T-t) }\le C(T-t),
\end{align*}
because the derivative $(\beta^{(i)})'$ is bounded on $[0,T]$ (the constant $C$ does not depend on $T$ as long as $T< T_0^\text{Riccati}$). The estimate involving $\gamma^{(i)}$ is similar and \eqref{estimate3} follows.  

By combining the estimates \eqref{estimate1}, \eqref{estimate2}, and \eqref{estimate3} we produce
\begin{align}
&\abs{\partial_{y_d}u^{(i)}(t,y)-\partial_{y_d}\tilde{u}^{(i)}(t,y) } \nonumber\\
&\leq \abs{\partial_{y_d}u^{(i)}(t,y)-\partial_{y_d}g^{(i)}(y) } + \abs{\partial_{y_d}g^{(i)}(y) -\partial_{y_d}\tilde{g}^{(i)}(y) }
+\abs{\partial_{y_d}\tilde{g}^{(i)}(y)- \partial_{y_d}\tilde{u}^{(i)}(t,y) } \nonumber\\
&\leq  C \Big((T-t)^{\frac{1+\alpha}{2}} + |y|^{1+\alpha} + (T-t)(1+\abs{y})\Big)\label{estimate4}.
\end{align}

Finally, by taking expectation through \eqref{estimate4} we obtain
\begin{displaymath}
\begin{split}
\mathbb{E}\Big[& \Big\vert \partial_{y_d}u^{(i)}(t,Y_t)-\partial_{y_d}\tilde{u}^{(i)}(t,Y_t) \Big\vert \Big] \\
&\leq  \int_{\R^D} \Gamma(0,t,y) \, \big\vert \partial_{y_d}u^{(i)}(t,y)-\partial_{y_d}\tilde{u}^{(i)}(t,y) \big\vert dy \\
&\leq C\int_{\R^D} \Gamma(0,t,y)  \Big(T^{\frac{1+\alpha}{2}} + \abs{y}^{1+\alpha} + T(1+\abs{y})\Big)dy \leq CT^{\frac{1+\alpha}{2}}.
\end{split}
\end{displaymath}
The last inequality holds because we are considering $T\in (0,1]$ and since
\begin{displaymath}
\begin{split}
\int_{\R^D} \Gamma(0,t,y) \abs{y}^{1+\alpha} dy &\leq c\int_{\R^D}\frac1{t^{D/2}} e^{-\frac{|y|^2}{\od t}}\abs{y}^{1+\alpha} dy \leq c T^\frac{1+\alpha}{2},
\end{split}
\end{displaymath}
for $t\in[0,T]$. This estimate follows from the definition \eqref{Gamma} of $\Gamma$ and the bounds provided in Lemma \ref{lem:LA}

$\endproof$


\begin{thebibliography}{}


\bibitem{BS12} S. Biagini and M. Sirbu (2012): \emph{
A note on admissibility when the credit line is infinite},
Stochastics, {\bf 84}, 157--169.

\bibitem{CLM12} P. O. Christensen, K. Larsen, and C. Munk (2012): \emph{Equilibrium in securities markets with heterogeneous
investors and unspanned income risk}, J. Econom. Theory \textbf{147},  1035--1063

\bibitem{CL12} P. O. Christensen and K. Larsen (2012): \emph{Incomplete continuous-time securities markets with stochastic income volatility}, Working paper, \url{http://www.andrew.cmu.edu/user/kasperl/}.

\bibitem{CH94} D. Cuoco and H. He (1994): \emph{Dynamic equilibrium in infinite-dimensional economies with incomplete financial markets}, Working paper.

\bibitem{DJ03} R. Dana and  M. Jeanblanc (2003): \emph{Financial markets in continuous time}, Springer.

\bibitem{Duf01} D. Duffie (2001): \emph{Dynamical asset pricing theory}, 3rd Ed., Princeton University Press.

\bibitem{DK96} D. Duffie and R. Kan (1996): \emph{
A yield-factor model of interest rates},
Mathematical Finance, {\bf 6}, 379--406.


\bibitem{Eva10} L. C. Evans (2010): \emph{Partial differential equations}, 2rd Ed., American Mathematical Society.

\bibitem{Fri64} A. Friedman (1964): \emph{Partial differential equations of parabolic type},  Prentice-Hall Inc.

\bibitem{HJ13} R. A. Horn and C. R. Johnson (2013): \emph{Matrix Analysis}, 2nd Ed., Cambridge University Press.

\bibitem{KS91} I. Karatzas and S. E. Shreve (1991): \emph{Brownian motion and stochastic calculus}, 2nd Ed., Springer.
 
\bibitem{KS98} I. Karatzas and S. E. Shreve (1998): \emph{Methods of mathematical finance}, Springer.

\bibitem{Kry96} N.V. Krylov (1996): \emph{Lectures on Elliptic and Parabolic Equations in Hšlder Spaces}, Graduate Studies in Mathematics, vol. 12. AMS, Providence.

\bibitem{LSU68} O. A. Ladyzenskaja, V. A. Solonnikov, and N. N. Ural'ceva  (1968): \emph{Linear and quasilinear equations of parabolic
type}, Translations of Mathematical Monographs, vol. 23. AMS, Providence.

\bibitem{SC02} A.N. Shiryaev and A.S. Cherny (2002): \emph{
Vector stochastic integrals and the fundamental theorems of asset pricing},
Proceedings of the Steklov Mathematical Institute, {\bf 237}, p. 12-56.

\bibitem{Zha12} Y. Zhao (2012): \emph{Stochastic equilibria in a general class of incomplete Brownian market environments}, Ph.D. thesis from UT-Austin.

\bibitem{Zit12} G. {\v Z}itkovi\'c (2012): \emph{An example of a stochastic equilibrium with incomplete markets}, Finan. Stoch. {\bf16},
177--206.

\end{thebibliography}
\end{document}